\documentclass[fleqn,usenatbib]{mnras}
\usepackage[dvipsnames]{xcolor}
\usepackage{newtxtext,newtxmath}
\usepackage[T1]{fontenc}
\DeclareRobustCommand{\VAN}[3]{#2}
\let\VANthebibliography\thebibliography
\def\thebibliography{\DeclareRobustCommand{\VAN}[3]{##3}\VANthebibliography}

\usepackage{graphicx}	
\usepackage{amsmath}	

\title[SGRB jets in structured and expanding ejecta]{Effects of polar structure and moving ejecta on the dynamics of SGRB jets}

\author[Garc\'ia-Garc\'ia et al.]{Leonardo Garc\'ia-Garc\'ia$^{1}$\thanks{E-mail: lgarcia@astro.unam.mx}, Diego L\'opez-C\'amara$^{2,3}$, Davide Lazzati$^{4}$
\\
$^{1}$Instituto de Astronomía, Universidad Nacional Autónoma de México, CDMX  04510, M\'exico\\
$^{2}$Instituto de Ciencias Nucleares, Universidad Nacional Aut{\'o}noma de M{\'e}xico, A. P. 70-543 04510 D. F. Mexico \\
$^{3}$Investigador por M\'exico, CONAHCyT -- Universidad Nacional Aut\'onoma de M\'exico, Instituto de Astronom\'ia, AP 70-264, CDMX  04510, M\'exico\\
$^{4}$Department of Physics, Oregon State University, 301 Weniger Hall, Corvallis, OR 97331, U.S.A.}

\date{Accepted XXX. Received YYY; in original form ZZZ}

\pubyear{\the\year{}}

\begin{document}
\label{firstpage}
\pagerange{\pageref{firstpage}--\pageref{lastpage}}
\maketitle

\begin{abstract}
At least some short gamma-ray bursts (SGRBs) originate from neutron star mergers, systems that release both a relativistic collimated jet and slower, wider ejecta. These jets evolve \textcolor{black}{through a dense, anisotropic, and expanding medium} produced during the merger process, resulting in interactions that affect their morphology and observable signatures. We investigate the propagation of SGRB jets through funnel-like structures that can be static or expanding with mildly relativistic speed using 2D axisymmetric relativistic hydrodynamic simulations. \textcolor{black}{Our initial conditions are inspired from radial and angular distributions of density and pressure from general-relativistic magnetohydrodynamic} simulations of binary neutron star mergers. We explore different values of the funnel opening angle and density contrast. We find that the polar structure of the ejecta mainly affects the jet evolution in the early stages, whereas the effect of expanding ejecta dominates in later stages. Jets propagating through a low-density polar funnel move initially faster, while the presence of mildly relativistic ejecta maintains the outflow more collimated after the breakout. \textcolor{black}{Despite differences in the external medium, the energy dissipation within the jet and cocoon remains similar across models, while the shocked ambient material shows distinct signatures that could be observationally distinguishable.} Our results highlight the importance of the structure and dynamical properties of the ejecta in shaping SGRB jets.
\end{abstract}


\section{Introduction}
Short gamma-ray bursts (SGRBs) are intense, transient events with characteristic gamma-ray isotropic energies in $\sim10^{49}-10^{52}$~erg \citep{Berger2014}. \textcolor{black}{At least some of them originate from the coalescence of two neutron stars (NSs)  \citep[which our study will be based upon, e.g.,][]{Goodman1986, Paczynski1986, Eichler1989, Narayan1992, Abbot2017A}. Other SGRB progenitor scenarios include the merger of a NS with a black hole (BH)  \citep[e.g.,][]{Paczynski1991, Narayan1992, Lee1995, Kluzniak1998, Janka1999}. For a review of the characteristics and progenitors of SGRBs, see \citep{Berger2014}}. After the merger, a relativistic outflow with an energy of approximately $10^{51}$~erg is released with a typical duration of $t_{90}\leq 2$~s \citep{Kouveliotou1993}. 
The jets can achieve Lorentz factors as high $\Gamma\sim10^{3}$ \citep{Ghirlanda2018} and isotropic luminosities between $L_{\rm{iso}}\sim10^{50}-10^{53}$~erg~s$^{-1}$ \citep{Ghirlanda2009}. These jets exhibit opening angles of approximately $\theta_{\rm{j}}\sim 0.5^{\circ}-25^{\circ}$ \citep{rouco2022}, resulting in intrinsic luminosities of $L_{\rm{j}} \sim 10^{45}-10^{51}$~erg~s$^{-1}$. 

\textcolor{black}{An important dynamical process in these systems is the interaction of the relativistic jet with the dynamical ejecta (produced and launched during the merger of the NSs) \citep{Murguia-Berthier2021, Urrutia2021, Salafia2022}.}
According to observations and general relativistic magnetohydrodynamic (GRMHD) simulations, the ejecta have a mass of $\sim 10^{-3}-10^{-2} M_{\odot}$ \citep{Cowperthwaite2017, Smartt2017, Tanaka2017, Ruiz2021}, densities of order  $\sim (10^{8}-10^{14})$ g~cm$^{-3}$, and magnetic field strengths of $\sim (10^{12}-10^{16})$~G \citep{Ciolfi2017, Ruiz2021, Combi2023}. In addition, dynamical ejecta have expanding velocities of the order of $(0.1-0.3)\rm{c}$ \citep{Cowperthwaite2017, Smartt2017, Kasliwal2017}, with a fraction reaching velocities exceeding $>0.4c$ \citep{Hotokezaka2013, Rosswog2024}. \textcolor{black}{Throughout this work, we refer to the expanding ejecta component as the wind. This interaction creates a complex network of shocks that greatly influences the morphology of the jet and the resulting radiation \citep[e.g., ][]{Gottlieb2021c, Salafia2022, Mpisketzis2024}.} 

\textcolor{black}{SGRB modeling has examined the propagation of relativistic and collimated outflows through environments with masses on the order of $\sim 10^{-2} M_{\odot}$} through both analytical  \textcolor{black}{\citep[e.g.,][]{LazzatiPerna, Hamidani2021, GG2024, Hamidani2024} and numerical studies \citep[e.g.,][]{Lazzati2017, Gottlieb2018a, Gottlieb2018b, Hamidani2020, Gottlieb2021b, Urrutia2021, GG2023}.}
\textcolor{black}{Increasing attention has been paid to modeling anisotropic post-merger environments \citep[e.g.,][]{Geng2019, Pavan2021, Pavan2023, Pavan2025, Nativi2022, Nativi2023, Mattia2024, Mpisketzis2024, Pais2024}. 
Recent GRMHD simulations have shown that, during the merger of two compact objects, the combined action of gravitational, centrifugal, and pressure gradients generates a non-uniform environment, often featuring low-density polar regions commonly referred to as “funnels” \textcolor{black}{\citep{Rosswog2000,  Pavan2021}}. These structures may improve the jet breakout efficiency by reducing the jet's baryon loading and energy loss during its propagation through the ejecta \citep{Pavan2021}. Neutrino winds \textcolor{black}{may be present \citep[e.g.][]{Aloy2005, Berthier2017, Nativi2021}} and may also develop an angularly stratified environment with low-density polar regions due to disk outflows \citep[][]{Perego2014, Sekiguchi2016,  Hayashi2022}. Both funnel configurations (that shaped by gravitational forces or from neutrino-driven winds) may share a similar morphology, but differ in their origin, composition, and dynamical evolution.}

\citet{LazzatiPerna} examined analytically the effect that a uniform and expanding wind has on the evolution of a SGRB-like jet (showing how the expansion of the medium influences the energy profile).
\citet{Geng2019} examined through 2.5D relativistic magnetohydrodynamic (RMHD) simulations the time delay influence of an expanding medium with a funnel in the evolution of relativistic jets, demonstrating its impact on jet structure and angular distribution.
\citet{Hamidani2021} analyzed the propagation of an SGRB jet through analytical and semi-analytical models (calibrated with a set of 2D RHD simulations), to study its interaction with an expanding wind without a funnel (as well as a long GRB in the Collapsar scenario), showing how the density and velocity profiles of the medium can influence the morphology of the jet.
\citet{Gottlieb2022} provides information on the evolution of a magnetized jet propagating through a homologous expanding medium, exploring the evolution and overall dynamics analytically and via three-dimensional RMHD simulations. \textcolor{black}{\citet{Pavan2021} investigated the propagation of a relativistic jet through BNS merger ejecta, where the ejecta structure was pre-computed using a GRMHD simulation. Their analysis revealed that a pre-existing low-density funnel improves jet breakout efficiency. This funnel structure enables the jet to maintain higher energy by reducing baryon loading and interaction losses during its propagation through the ejecta.}

\citet{Pais2023, Pais2024} analyze GRMHD simulations and investigate the effects that a structured and expanding medium has on the dynamics of different SGRB-like jets. The jets vary their luminosity and opening angle while the medium has a fixed funnel. Their findings revealed that the interaction of the medium with the jet influences its dynamics and modifies its structure.

This study aims to investigate the impact of different structured and expanding media on the evolution of a relativistic jet. In order to provide a comprehensive understanding, we study the global effects and also analyze the individual contributions that the structure and expansion of the medium produce. 
This paper is organized as follows. Section~\ref{s:setup} describes the setup and physics in our simulations. The results of the propagation of a relativistic jet through a funnel medium are presented in Section~\ref{s:results}. Finally, Section~\ref{s:D&C} presents the summary and conclusions.

\section{Global setup and models}
\label{s:setup}
In order to study the evolution of an SGRB-like jet through different structured and static or expanding media, \textcolor{black}{we perform a set of two-dimensional axisymmetric RHD simulations using the PLUTO code (v4.4; \citealt{Mignone2007}) in spherical coordinates. The simulations employ a Harten–Lax–van Leer \citep[HLL;][]{HLL} Riemann solver and a third-order Runge–Kutta time integration scheme \textcolor{black}{(for details see \citep{gottlieb1996} and \citep{Mignone2007})}.We model the gas evolution assuming a relativistic equation of state with an adiabatic index of 4/3.}

\begin{figure}
    \centering
    \includegraphics[width=\columnwidth]{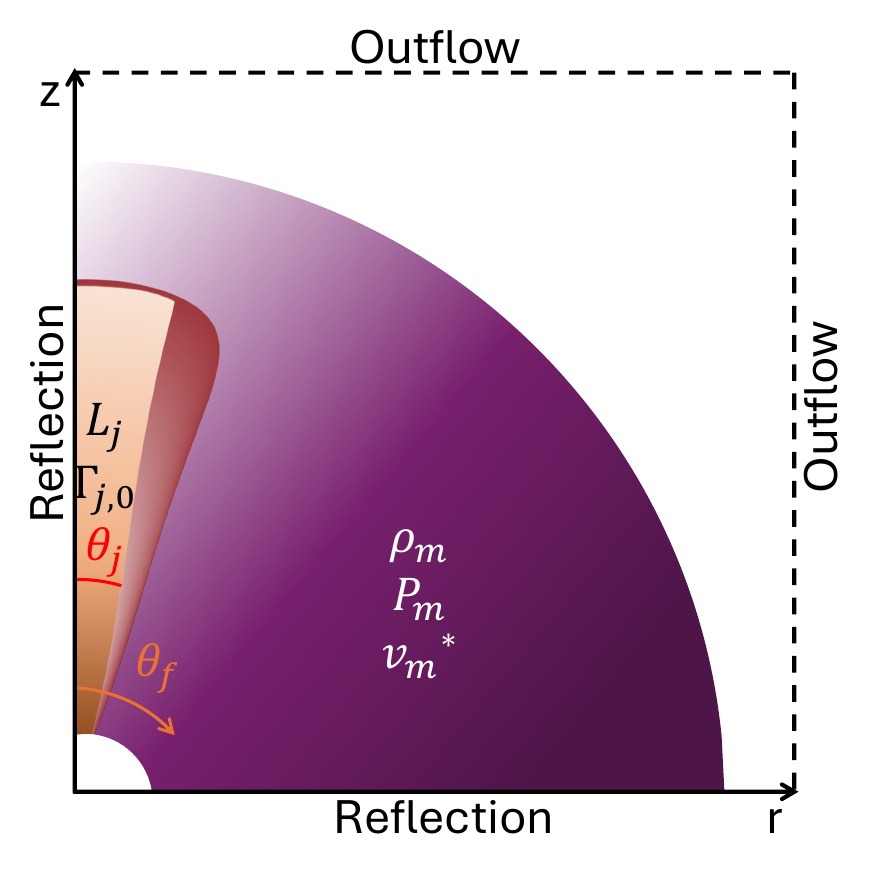}
    \caption{Setup scheme showing a relativistic jet (orange) and its cocoon (red) evolving through a medium (purple) with a funnel (white) with opening angle $\theta_{\rm{f}}$.
    The jet was launched from an inner boundary ($r_{\rm{min}}$) with a luminosity $L_{\rm{j}}$, initial Lorentz factor $\Gamma_{\rm{j,0}}$, and opening angle $\theta_{\rm{j}}$. The medium, with density $\rho_{\rm{m}}$ and pressure $P_{\rm{m}}$, may have an angular structure including a low-density funnel with an opening angle $\theta_{\rm{f}}$ and may be static or may expand with constant velocity $v_{\rm{m}}$. The boundary conditions are also shown.}
    \label{f:setup}
\end{figure}

Figure \ref{f:setup} illustrates the setup used in our simulations. A relativistic and collimated jet, characterized by its luminosity $L_{\rm{j}}$, opening angle $\theta_{\rm{j}}$, and initial Lorentz factor $\Gamma_{\rm{j,0}}$, propagates through a stratified medium defined by its density $\rho_{\rm{m}}$ and pressure $P_{\rm{m}}$. Due to the modeling in 2D, the jet is axisymmetric. The medium may have an angular structure, including a low-density funnel with an opening angle $\theta_{\rm{f}}$, which provides a channel for the expansion of the jet. \textcolor{black}{Depending on the configuration, the medium can either be approximated as static (if the jet velocity significantly exceeds the medium's expansion speed) or expand outward with a constant velocity $v_{\rm{w}}$ (since the dynamical ejecta is expelled at mildly relativistic velocities shortly before the merger).}

The density of the medium ($\rho_{\rm{m}}$) is composed of four components and is shown in Equation~\ref{eq:density_m}. \textcolor{black}{Based on the GRMHD simulations of \citet{Ciolfi2017} (with equal NS masses and a soft equation of state, model APR4), and as in \citet{GG2023}, the density profile of the medium was found to follow a $\rho(r)=\rho_0 (r/r_0)^{-3}$ radial profile\footnote{\textcolor{black}{The profile was obtained by averaging multiple radial profiles sampled at various polar angles from the 2D meridional density plane of the equal-mass APR4 model presented in \citet{Ciolfi2017}. The resulting profile was smoothed, and a $\rho \propto r^{-n}$ profile with an index of $n = 2.97 \approx 3$ was obtained.}} (with $r_0 = 1.5\times10^{7}$~cm and $\rho_{0}=3.1\times10^{8}\rm{g}~\rm{cm}^{-3}$).}

\textcolor{black}{An exponential cut-off point, adapted from \citet{Lazzati2017}) was included and set to $r_{\rm{cut}} =5\times10^{8}$~cm in order for the medium to have a total mass of $1.1 \times 10^{-2} M_{\odot}$.} \textcolor{black}{We follow the approach of \citet{Kawaguchi2020A} to introduce the angular stratification of the environment (where the opening angle of the funnel is set by $\theta_{\rm f}$). We add the constant $\eta$ to define the density ratio between the funnel ($0 \leq \theta \leq \theta_{\rm f}$) and the equator ($\theta_{\rm f} \leq \theta \leq \pi/2$). For $\eta = 1$ there is no density difference between the funnel and the equator, while for $\eta = 9$ the funnel is nine times less dense than the equator.}
Finally, a background density\footnote{\textcolor{black}{The background density is not taken into account when calculating the mass of the medium.}} ($\rho_{\rm{b}}=3\times10^{-3} \rm{g}~\rm{cm}^{-3}$) was added. \textcolor{black}{The density of the medium $\rho_{\rm{m}}$ reads:}

\begin{equation}
\label{eq:density_m}
   \rho_{m}(r, \theta) = \frac{\rho_{0}\left(\frac{r}{r_{0}}\right)^{-3} e^{\left(-\frac{r}{r_{cut}}\right)^3}}{1+\left(\eta-1\right)e^{ -10\left(\frac{\theta}{\theta_{f}}\right)^3 }} + \rho_{b}.
\end{equation}
\textcolor{black}{The initial density profile for a static medium with a funnel with $\eta = 6$ and $\theta_f = 30^{\circ}$ is shown in Appendix \ref{s:initial_cond}.}
\textcolor{black}{The pressure of the medium is taken to be an ideal gas (with an adiabatic index of $\gamma=5/3$) with a temperature of $T =10^{7}$~K. The medium may be static or may be expanding radially with $v=0.1~c$.}

\textcolor{black}{We manually inject a SGRB-like jet by imposing boundary conditions at the inner boundary radius $r = r_{\rm{min}}$. 
The jet has a constant luminosity of $L_{\rm{j}}=10^{51}$~erg~s$^{-1}$, half opening angle $\theta_{\rm{j}} = 5^{\circ}$, an initial Lorentz factor $\Gamma_{\rm{j,0}}=5$ and the Lorentz factor at infinity of $\Gamma_{\rm{\infty}}=400$. The jet properties are derived using a prescription similar to \citet{Pavan2023} (neglecting the magnetic fields and the rest mass). }
\textcolor{black}{From the relativistic stress-energy tensor for a fluid in motion, the luminosity is $L=\int_\Sigma T^{0i} d\Sigma$. Thus, the jet density ($\rho_j$) at the base of the jet is:
\begin{equation}
 \rho_{\rm{j}} = \frac{L_{\rm{j}} }{\pi c^{2} r_{\rm{ap}}^2 v_{\rm{j}}  \Gamma_{j,0} \Gamma_{\infty} }
\end{equation}}
\textcolor{black}{where $r_{\rm{ap}}$ is the aperture radius ($r_{\rm{ap}}=r_{\rm{min}}tan\theta_{\rm{j}}$) and $v_{\rm{j}}$ is the jet velocity.
Meanwhile, the jet pressure ($p_{\rm{j}}$) is:
\begin{equation}
p_{\rm{j}} = \frac{\rho_{\rm{j}} c^2}{4} \left( \frac{\Gamma_{\infty}}{\Gamma_{\rm{j,0}}} - 1 \right).
\end{equation}}
The medium is modeled with varying funnel configurations, characterized by different opening angles $\theta_{\rm{f}}$ and density contrasts $\eta$ \textcolor{black}{(see Figure~\ref{f:init_cond} for one of the density initial conditions that we used).}

\textcolor{black}{While motionless environments do not fully capture post-merger conditions, they provide a useful reference to isolate the role of the funnel. Thus, we first explored the effects produced by the funnel; then we examined its interaction with an environment expanding with $v_{\rm{w}} = 0.1\,c$. We consider two funnel opening angles, $\theta_{\rm{f}} = 15^{\circ}$ and $30^{\circ}$, along with density contrast values of $\eta = 1, 3, 6, 9$.}

The models are named according to their specifics. The acronym ``nF'' signifies the absence of a funnel, while ``F'' denotes the presence of a funnel. Similarly, nW'' represents a static medium, while W indicates an expanding wind. Furthermore, ``l'' corresponds to a case in which the funnel is less dense (light, $\eta=9$), and ``h'' indicates that the funnel was denser (heavy, $\eta=3$), both compared to their standard funnel correspondent case (F\_nW and F\_W). The narrow funnel case ($\theta_{\rm{f}}=15^{\circ}$) is denoted by ``N''. Table~\ref{t:table} summarizes the parameters used in each model.

\begin{table}
\centering
\caption{\textcolor{black}{Summary of model parameters. Each model is characterized by the funnel opening angle ($\theta_{\rm f}$), the density contrast between the polar and equatorial regions ($\eta$), and the expanding velocity of the environment ($v_{\rm w}$). Models labeled with ``W'' include an expanding wind component, while ``nW'' indicates a static medium.}}

\begin{tabular}{cccc}
\hline
Model & $\theta_{\rm{f}}$ ($^\circ$) &  $\eta$ & $v_{\rm{w}} / \rm{c}$  \\
\hline
\hline
nF\_nW  & 0   & 1  & 0  \\
Fl\_nW  & 30  & 3  & 0  \\
F\_nW  & 30  & 6  & 0  \\
Fh\_nW  & 30  & 9  & 0  \\
FN\_nW  & 15  & 6  & 0  \\
nF\_W  & 0   & 1  & 0.1 \\
F\_W  & 30  & 6  & 0.1 \\
\hline
\end{tabular}
\label{t:table}
\end{table}

Except for the jet injection region, at $r=r_{\rm{min}}$ and between $\theta_{\rm{min}}=0^{\circ}$ and $\theta=\theta_{\rm{j}}=5^{\circ}$, the inner boundary had a reflective condition. The r$_{\rm{max}}$ boundary had a free outflow condition. The axisymmetric and equatorial boundaries ($\theta_{\rm{min}}$ and $\theta_{\rm{max}}$, respectively) had reflective conditions. The simulations were carried out with a Courant number of $\rm{Co}=0.3$, \textcolor{black}{and the total integration time was limited to $t = 0.124$~s due to computational constraints.} The computational domain extended from $r_{\rm{min}}=2\times10^{7}$~cm to $r_{\rm{max}}=3.62\times10^{9}$~cm and between $\theta_{\rm{min}}=0$ and $\theta_{\rm{max}}=\pi/2$. The resolution comprised $N_{r}=30,000$ cells with a logarithmically increasing radial spacing and $N_{\theta}=1,000$ uniformly sized angular cells. \textcolor{black}{We extend the computational domain and maintain the spatial resolution employed in \citet{GG2023} where a comprehensive convergence study was conducted.}

\section{Results }
\label{s:results}

In this section, we describe the results of our simulations. We focus on overall differences driven by the presence or absence of characteristic components (funnel and wind) in each simulation. The implications of these differences for potential observational effects are discussed in Section~\ref{s:D&C}.

\subsection{Funnel and wind effects}
\begin{figure*}
    \centering
    \includegraphics[width=1.99\columnwidth]{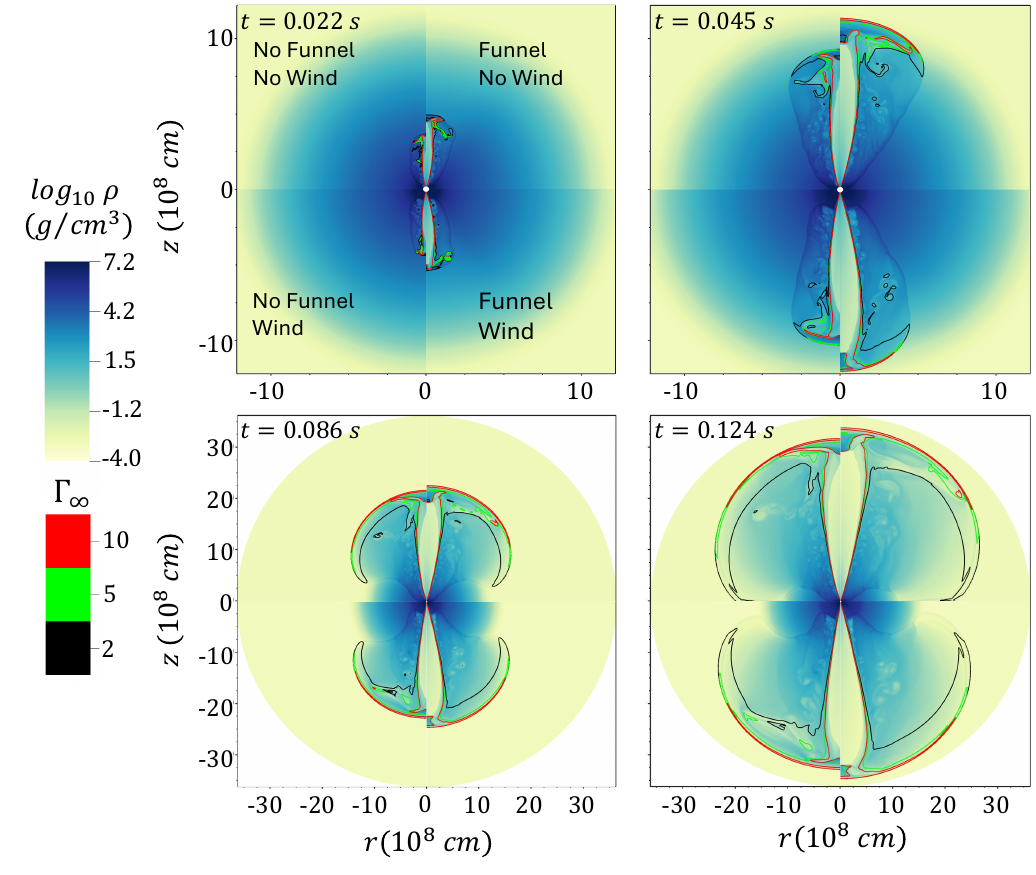}
    \caption{Density maps ($\rm{g} \ \mathrm{cm}^{-3}$) for a relativistic jet moving through a static medium without funnel (nF\_nW, upper left), the static medium with funnel (F\_nW, upper right), the expanding medium without funnel (nF\_W, lower left), and the expanding medium with funnel (F\_W, lower right). Four different times are shown ($\rm{t} = 0.022~\mathrm{s}$, $0.045~\mathrm{s}$, $0.086~\mathrm{s}$, and $0.124~\mathrm{s}$). Isocontours of $\Gamma_{\infty} = 2$, $5$, and $10$ are depicted in black, green, and red, respectively. Note that the spatial domain sizes differ between the upper and lower panels.}
    \label{f:dens_plot}
\end{figure*}

Figure~\ref{f:dens_plot} presents density maps for the evolution of a SGRB-like jet under four scenarios: a static medium without a funnel (control case, nF\_nW, upper left), a static medium with a funnel (F\_nW, upper right), an expanding medium without a funnel (nF\_W, lower left), and an expanding medium with a funnel (F\_W, lower right). The evolution of the jet is presented at four different times $\rm{t} = 0.022~\mathrm{s}$, $0.045~\mathrm{s}$, $0.086~\mathrm{s}$, and $0.124~\mathrm{s}$ (corresponding to the maximum integration time). The isocontours of $\Gamma_{\infty} = 2$, $5$, and $10$  are overlaid in black, green, and red, respectively.

A first general consideration that we find is that the presence of a funnel has the strongest impact on early jet evolution, while the wind mostly affects the large-scale behavior of the outflow. Consider, for example, the case in which the jet is in its early stage, at $\rm{t} = 0.022~\mathrm{s}$, drilling through the different media (shown in the upper left panel of Figure~\ref{f:dens_plot}). There is a marked difference between the left and right sides of the upper left panel (that is, the cases in which the medium does not have and has a funnel, respectively). The top and bottom parts, instead, are similar, showing that the presence or absence of a wind is barely noticeable at these times. 
Instead, as time progresses, the presence of a wind becomes more noticeable. For the case in which $\rm{t} = 0.124~\mathrm{s}$, there is a clear difference between the top and bottom parts of the bottom right panel of Figure~\ref{f:dens_plot} (i.e., those with a static and expanding medium, respectively). The static case is characterized by extended angular wings that reach all the way to $90^{\circ}$. The presence of wind, instead (bottom part of the panel), produces hydrodynamical collimation at a large distance, resulting in a more focused outflow. We now examine each case in more detail and quantify the differences.

Jets evolving through a medium that has a funnel evolve faster $\approx 10\%$ faster) compared to the case in which there is no funnel. In addition, the funnel produces a larger jet opening angle. \textcolor{black}{The latter two funnel effects are visible in Figure \ref{f:jet-head}, where the evolution of the jet head over time for various models is shown.}  \textcolor{black}{Meanwhile, jets that evolve through expanding media take more time to break out (see the dots and triangles in Figure \ref{f:jet-head})}. 

In the breakout, the opening angle for the material with $\Gamma = 2$ is $6^{\circ}$ is wider in the static medium with a funnel (F\_nW) while $6^{\circ}$ narrower in the expanding wind without a funnel (F\_nW), compared to the static medium without a funnel (nF\_nW, for which $\theta \approx 29^\circ$). 
After the breakout, the combination of the funnel and the expanding wind produces a narrower opening angle of the jet head.
The material with $\Gamma = 2$, in the breakout (and final integration time), had $\theta \approx 29 ^{\circ}$ ($\theta \approx 75^{\circ}$) and $\theta \approx 33^{\circ}$ ($\theta \approx 90^{\circ}$) for models F\_W and nF\_nW, respectively.

Similar results to those of the standard funnel case shown in Figure~\ref{f:dens_plot} were observed for lighter and heavier funnels (in a static medium). Minor variations in the jet head velocity were found ($\lesssim 1$).  However, the lateral expansion exhibited differences of $1\%$ for Fl\_nW and $16\%$ for Fh\_nW compared to the F\_nW model. The narrower funnel case produced a jet $3.4\%$ slower than the F\_nW model with fewer lateral expansion. The lateral expansion for $\bar{\Gamma}_{\infty} =2$ of the narrow funnel was $\sim 10^\circ$ more acute compared to the standard funnel case ($\theta_{\mathrm{f,FN\_nW}}\sim 25^{\circ}$ and $\sim 80^{\circ}$ before and after the jet breakout).

An analysis of the thermal and kinetic energy reveals that the material within the cocoon consists of both components \textcolor{black}{(see for example Figure~\ref{f:Kinetic_Energy} where the spatial and temporal evolution of the energy ratio, $e_k/e_{th}$, for various models is shown)}. As the jet evolves, substantial thermal energy conversion into kinetic energy occurs, particularly in the outer cocoon layers. Following the breakout, the cocoon transitions to being primarily kinetically dominated in both models, with regions of enhanced kinetic energy concentrated near the jet core. In addition, the fact that the cocoon is kinetically dominated agrees with the results from \citet{Hamidani2021}.

\subsection{Jet structure}
One aspect of our results that is potentially important for observations is the angular distribution of the jet energy and speed (or Lorentz factor). In this section, we take a deeper dive into this aspect of our results, focusing on the late-time properties that are the best proxy for the jet structure at the radiation stage.

\begin{figure}
    \centering
    \includegraphics[width=\columnwidth]{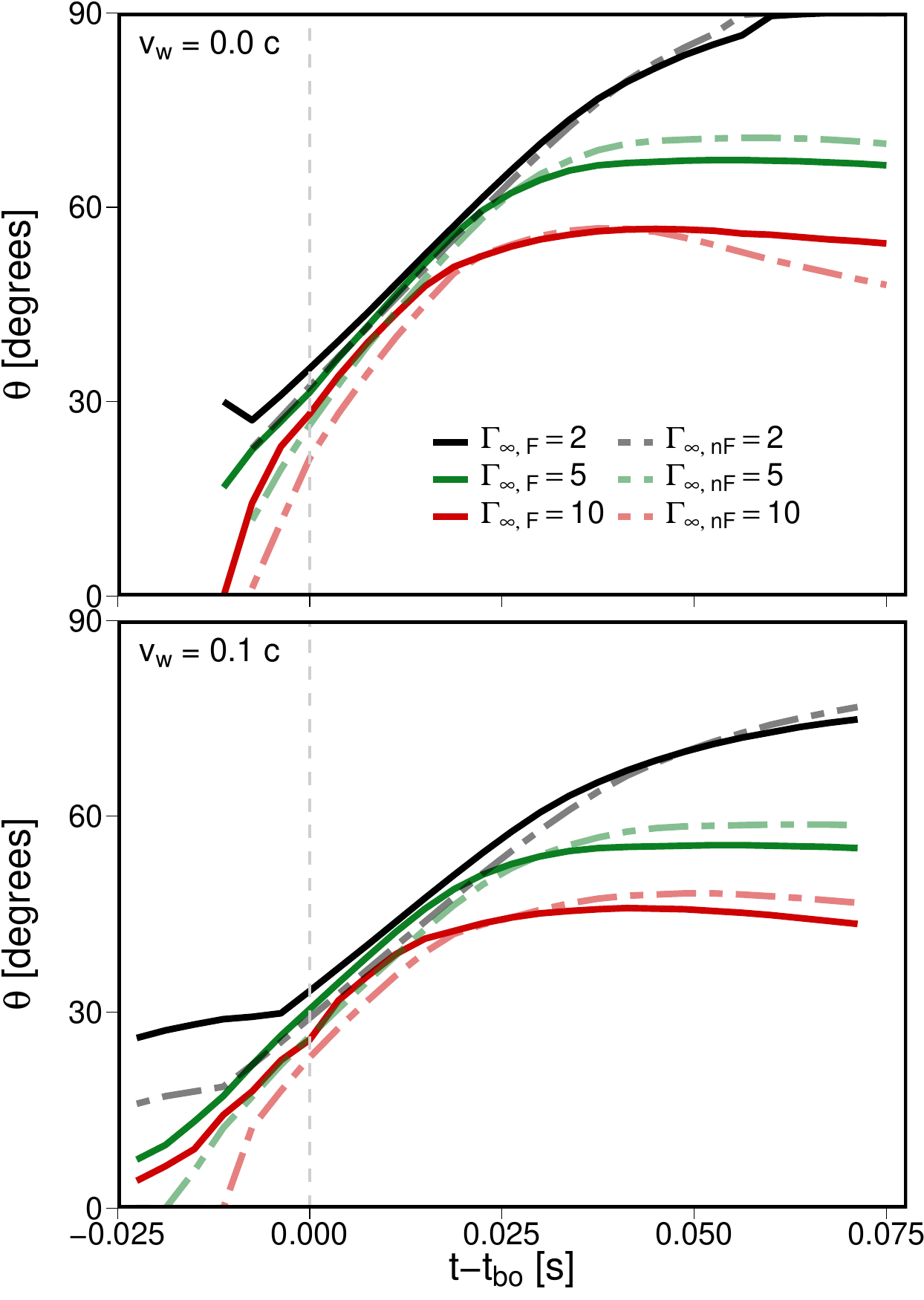}
    \caption{Maximum angle reached ($\theta$) by the material moving at different speeds over breakout time. Solid lines indicate funnel simulations, while dashed lines represent control simulations. Maximum angles for $\Gamma_{\infty} = 2$, 5, and 10 are marked in black, green, and red, respectively. Upper panels display F\_nW and nF\_nW simulations; lower panels show F\_W and nF\_W simulations.}
    \label{f:angles2}
\end{figure}

Figure~\ref{f:angles2} presents the opening angles of the fast-moving material in the jet-cocoon of our models as a function of time (normalized to their corresponding breakout time, $t-t_{\rm{bo}}$). 
\textcolor{black}{The opening angle of material moving at a certain Lorentz factor $\Gamma$ is defined as follows. We first select all the parcels of material that move faster than the selected value of $\Gamma$. For each of these parcels we know their off-axis angle as their polar coordinate. We select the one with the largest value of polar angle and define that angle as the angle $\theta$ that is plotted in Figure~\ref{f:angles2}.}
Different colors and line styles are used to identify the Lorentz factor considered and the possible presence of a funnel. The top panel shows the result for static media, while the bottom panel shows the results in a pre-accelerated medium. The vertical dashed gray line indicates the breakout time. The faster the jet material, the narrower the opening angles of the fast-moving material. For example, for the final integration time in our simulations, the opening angle for $\Gamma_{\infty} = 10$ is approximately half that of $\Gamma_{\infty} = 2$, independently of the characteristics of the medium. In addition, the opening angles of the jets are significantly narrower when propagating through an expanding medium. For example, the final opening angle in an expanding medium for $\Gamma_{\infty} = 2$ is always below $\sim 75^{\circ}$ while it reaches $90^{\circ}$  in a static medium ($\Gamma_{\infty} = 2, 5$ are reduced by $\sim 10^{\circ}$ and $\sim 5^{\circ}$, respectively).

\begin{figure}
    \centering
    \includegraphics[width=\columnwidth]{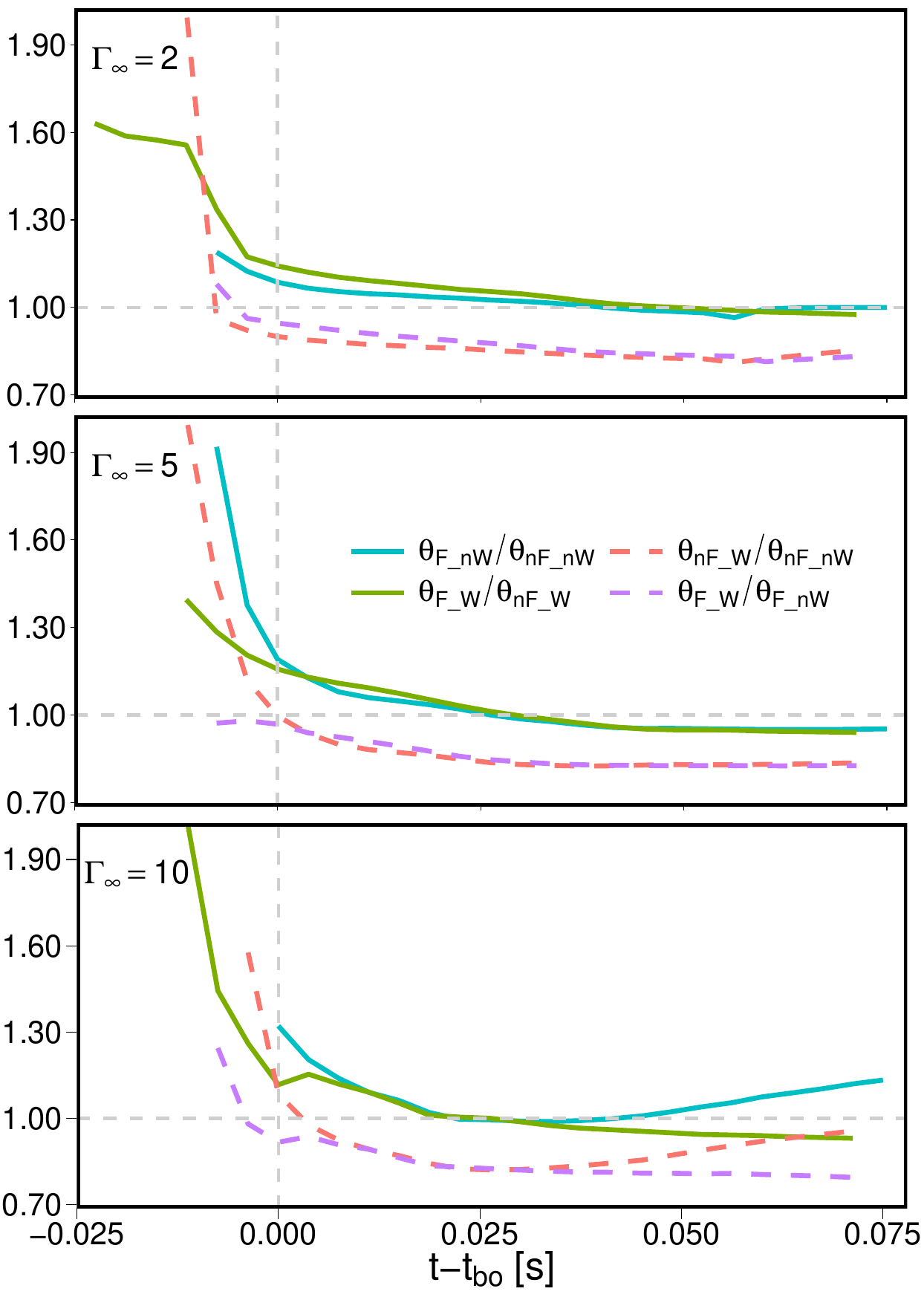}
    \caption{Ratio between angles from the models presented in Figure \ref{f:angles2} as a function of time $(t - t_{\text{bo}})$ for different values of $\Gamma_{\infty}$. The cyan lines represent the ratio $\theta_{\text{F\_nW}}/\theta_{\text{nF\_nW}}$, the green lines correspond to $\theta_{\text{F\_W}}/\theta_{\text{nF\_W}}$, the orange lines show $\theta_{\text{nF\_W}}/\theta_{\text{nF\_nW}}$, and the purple lines indicate the ratio $\theta_{\text{F\_W}}/\theta_{\text{F\_nW}}$. From top to bottom, the panels correspond to $\Gamma_{\infty} = 2$, 5, and 10.}
    \label{f:angles}
\end{figure}

To identify where the funnel or wind affects the evolution of the relativistic and collimated jet we plot in Figure~\ref{f:angles} the opening angle ratios $\theta_{\text{F\_nW}} / \theta_{\text{nF\_nW}}$ (cyan solid line), $\theta_{\text{F\_W}} / \theta_{\text{nF\_W}}$ (green solid line), $\theta_{\text{nF\_W}} / \theta_{\text{nF\_nW}}$ (orange dashed line) and $\theta_{\text{F\_W}} / \theta_{\text{F\_nW}}$ (purple dashed line) as a function of time. 
The first two ratios give insight into where the funnel effects are dominant, and the last two ratios indicate when the wind effects prevail. The panels, from top to bottom, correspond to $\Gamma_{\infty} = 2$, $\Gamma_{\infty} = 5$, and $\Gamma_{\infty} = 10$. 
Before the breakout time, the funnel effects dominated during the evolution of the jet and produced a broad jet. The funnel may increase the opening angle of the jet by up to twice compared to the case when no funnel is present (indicated by the upper left regions of the panels in Figure~\ref{f:angles}). On the other hand, after the breakout time, the wind effects become dominant and collimate the jet opening angle. In this phase, the opening angle of the jet is basically the same regardless of whether the medium had a funnel or not, and it may be reduced by $\sim 30 \%$ (compared to the case of static medium). For the funnel case with static medium (F\_nW, cyan line in the lower panel of Figure~\ref{f:angles}), the opening angle for the material with $\Gamma_{\infty} = 10$ is at all times larger than its non-funnel case.

\begin{figure}
    \centering
    \includegraphics[width=\columnwidth]{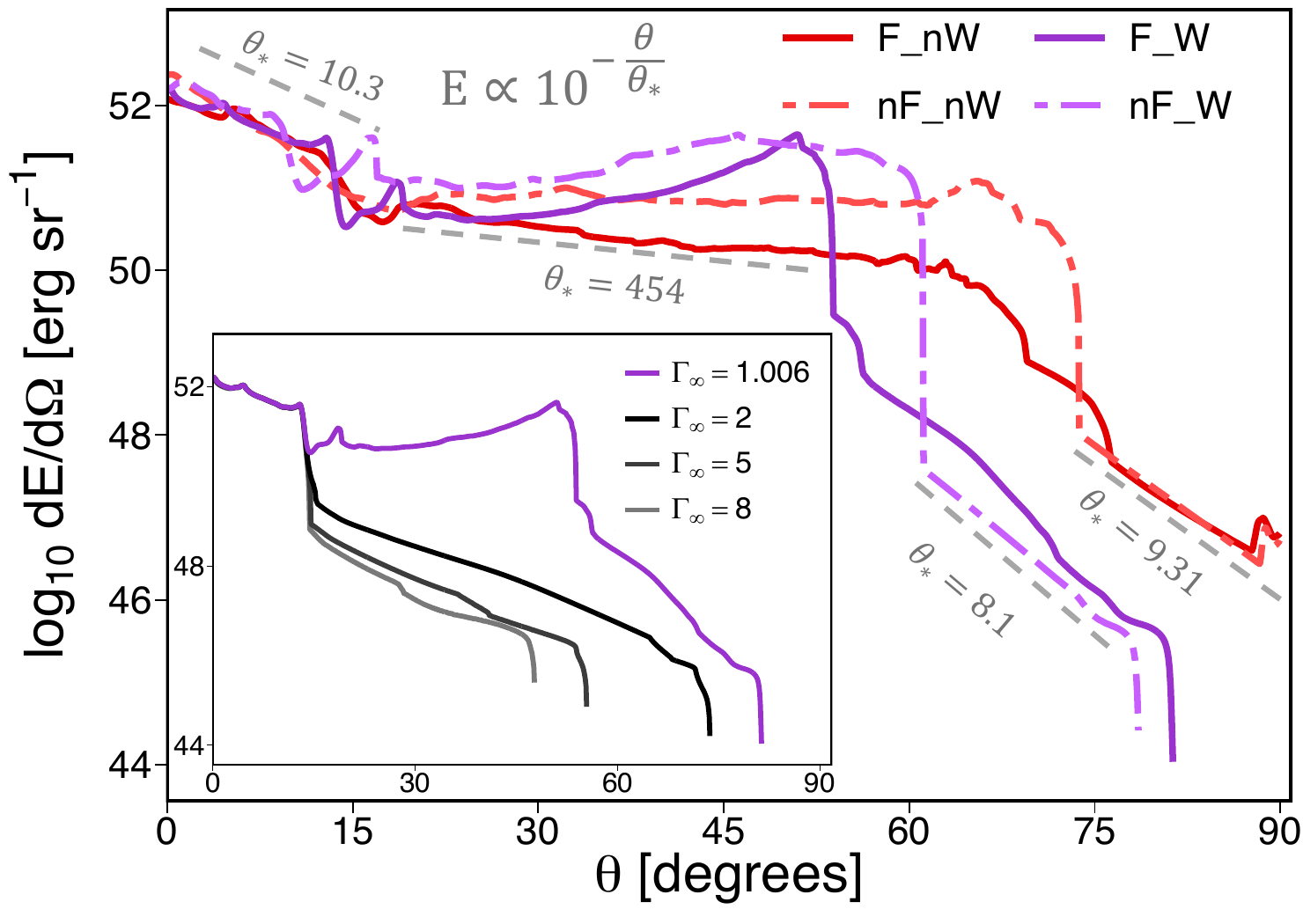}
    \caption{Energy per solid angle ($\mathrm{erg~sr}^{-1}$) profile as a function of angle for material with $\Gamma_{\infty}\geq1.006$ for models F\_nW  (red solid line), nF\_nW (red dot-dashed line), F\_W (purple solid line), and nF\_W (purple dot-dashed line). The gray dashed lines indicate different reference profiles. The inset figure presents the energy per solid angle profiles for model F\_W with $\Gamma_{\infty}\geq1.006, 2, 5, 8$ (purple, black, dark gray, and light gray lines, respectively).}
    \label{f:kinetc}
\end{figure}

\textcolor{black}{The energy per solid angle ($dE / d\Omega$) was computed from the relativistic stress-energy tensor for a fluid in motion. The Energy, $E$, was obtained using the expression $E = \int_\Sigma T^{00} d\Sigma$, following the formulation outlined by \citet{Landau1959}, \citet{Marti1999}, \citet{Mignone2007}, and \citet{Pavan2023}. This is:}
\begin{equation}
\label{eq:e_solid}
\textcolor{black}{\frac{dE}{d\Omega}~=~\int r^2 \left[ \Gamma^2 (\rho c^2 + 4p)  -p -\rho c^{2} \Gamma \right] \, dr}
\end{equation}
\textcolor{black}{where the rest mass is being removed and the density and pressure are a function of $r$ and $\theta$.} The integration limits go from $r_{\rm{min}}$ to the forward shock radius and the integral considers regions with $\Gamma_{\infty}\geq1.006$ (that is, $v \geq 0.11 c$).
Figure~\ref{f:kinetc} presents the $dE/d\Omega$ of our models (F\_nW, nF\_nW, F\_W, and nF\_W) for the final integration time. In addition , the inset figure presents the $dE/d\Omega$ for model F\_W considering $\Gamma_{\infty} \geq 1.006, 2, 5, 8$.
The slopes for three different regions are shown ($\propto 10^{-\theta / \theta_0}$, with $\theta_0$ a constant). Consistent with the findings of \citet{Bromberg2011, Lazzati2017}, the regions correspond to the jet and cocoon (containing jet material that has crossed the reverse shock) and an outer heavier part, also referred to as the external shock medium.
The jet is within $\sim 15^{\circ}$, reaches \textcolor{black}{$dE/d\Omega \sim 10^{51-52}$~erg~$\rm{sr}^{-1}$, and follows a profile with $\theta_0 = 10.3^{\circ}$}. 
The cocoon extends up to \textcolor{black}{$\sim 55^{\circ}-75^{\circ}$}, has a $\sim$plateau profile within \textcolor{black}{$10^{49-51}$~erg~$\rm{sr}^{-1}$}. For wind models, the plateau is followed by a sharp decay and vanishes if the material moves with high Lorentz factors ($\Gamma \geq 2$). The material within the jet has $dE /d\Omega \sim 10^{52}$~erg~sr$^{-1}$ independently of its Lorentz factor. Meanwhile, the material in the cocoon decreases its $dE /d\Omega$ as $\Gamma$ increases.
The rest of the material is the shocked medium \textcolor{black}{($\theta \geq 55^{\circ}-75^{\circ}$)}, has \textcolor{black}{$10^{44-49}$~erg~$\rm{sr}^{-1}$}. The static ejecta present more energetic shocked material (up to one or two orders of magnitude). For the static case \textcolor{black}{$\theta_0 =9.1^{\circ}$, while the expanding medium has $\theta_0 = 8.1^{\circ}$.}

\section{Summary \& discussion }
\label{s:D&C}
In this section, we summarize and discuss our results in the context of other studies in the field. In addition, we discuss what aspects are amenable to being tested with current or future observations.

This study explores the influence of two properties of the ambient material surrounding a BNS merger on the propagation of a relativistic outflow. These two properties are polar stratification (namely, the presence of a low-density funnel along the angular momentum direction) and the role of preexisting outward velocity. Our simulations demonstrate that a static funnel geometry enhances lateral cocoon expansion, increases the forward shock distance, and broadens the cocoon structure, especially in the early phase of the jet evolving through the media. The effect of the outward motion of the medium, instead, manifests itself at later times when a more effective collimation of the relativistic material is evidenced. Since the two explored components have an impact in different phases, it is not surprising that they are additive. When both are present, the effect is that of a funnel-driven early evolution coupled with increased collimation at later stages. 
\textcolor{black}{Additionally, the jets are collimated within the cocoon as the collimation condition, $\tilde{L} < \theta_0^{-4/3}$ (where $\tilde{L} = \rho_j h_j \Gamma_j / \rho_f$ and $\theta_j^{-4/3}$) of \citet{Bromberg2011}, is satisfied. Specifically, for the values in our models, we have $\tilde{L} \sim 2$ and $\theta_j^{-4/3} \sim 25$).}

Of the various effects discussed in this paper, the differences in angular energy profile have the most direct observational consequence. Figure~\ref{f:kinetc} shows that, independently of the presence of a funnel, an outflow propagating in a moving medium is characterized by much lower energy at viewing angles $\theta_v>60^{\circ}$. Although we cannot observe a single event at multiple viewing angles, it is anticipated that in the future we can compile a sample of bursts seen at different viewing angles, thanks to gravitational wave triggers, such as the case of GW170817. The ratio of bursts seen at low, intermediate, and large viewing angles would therefore allow us to infer and/or constrain the presence of moving ejecta.


\textcolor{black}{Our results for the expanding medium models show that the wind may affect the evolution of the jet and are consistent with the findings of \citet{Aloy2005}, who demonstrated that low-density polar funnels enhance jet acceleration, while \citet{Berthier2017} showed that winds can restrict lateral jet expansion. However, we obtain angular energy values that are $\sim$ three orders of magnitude larger (we notice that their jets are less energetic and that their ambient medium is different from ours, making a direct comparison not straightforward). Additionally, \citet{Nativi2021} further showed that asymmetric winds can effectively collimate the jet.}
In agreement with \citet{LazzatiPerna}, who investigated jets propagating through media with varying masses and velocities, we found that $\theta_{\rm{j}}$ and $\theta_{\rm{c}}$ are broader in lower-density environments.
The expanding wind with a funnel model produces angular energy distributions similar to those reported by \citet{Geng2019}, where the medium expands at $0.2\rm{c}$ and has a low-density polar structure ($\rho \propto \sin^3\theta$). Our energy profile shows a well-defined jet followed by a plateau if the cocoon evolves through an expanding medium. In contrast, they obtain narrower jets (by $\sim 5^{\circ}$) and their absence of data beyond $\theta \geq 50^\circ$ limits a comparison of the shocked material at large viewing angles.

\textcolor{black}{Our results are consistent with those of \citet{Pavan2021}, whose analysis demonstrates that a low-density funnel enhances jet breakout efficiency, enabling the jet to retain higher energy by reducing baryon loading and interaction losses during its propagation through the ejecta. This is consistent with our results, where we also observe that the presence of a low-density funnel prevents jet lateral expansion and boosts jet velocity.}
Our findings align with those of \citet{Pais2024}, which showed that a narrow funnel acts as a barrier leading to collimated outflows. In addition, we confirm that jets with high luminosities ($\sim 10^{51}$~erg~s$^{-1}$) lack recollimation shocks within the jet. To check that this is not a numerical effect or due to our numerical setup and geometry, we ran an additional simulation with lower jet luminosity ($5\times10^{50}$~erg~s$^{-1}$). We confirmed that recollimation shocks appear in the lower luminosity jet case.

We note that the present study is constrained to two-dimensional axisymmetric RHD simulations in a relatively small domain (at least one or two orders of magnitude short of the photospheric radius) and no radiation transport was included. The jet is taken to have a constant luminosity and its magnetization, as well as that of the medium, is not taken into account. In addition, the medium was modeled to expand with a constant velocity. Future investigations are required to explore the impact of these simplifications, including the extension to a larger three-dimensional computational domain, variable jet behavior, the role of magnetization, and the emission that would be produced.

\section*{Acknowledgements}
\textcolor{black}{We thank Fabio De Colle and Enrique Moreno Méndez for their valuable comments and suggestions which helped improve the quality of this work.} LGG and DLC acknowledge the support of the Miztli-UNAM supercomputer (project LANCAD-UNAM-DGTIC-321) and AWS (in the "Proyectos de investigación en la Nube UNAM-AWS") for the computational time assigned to perform the simulations and also acknowledge the support of the UNAM-PAPIIT grant IG100820.
DL acknowledges support from the NSF grant AST-1907955. \\
LGG acknowledges support from the CONAHCyT doctoral scholarship.\\
Many of the images in this study were produced using VisIt. VisIt is supported by the Department of Energy with funding from the Advanced Simulation and Computing Program, the Scientific Discovery through Advanced Computing Program, and the Exascale Computing Project.

\section*{Data Availability}
The data underlying this article will be shared on reasonable request with the corresponding author.

\bibliographystyle{mnras}
\bibliography{example}

\appendix
\section{Initial Density Distribution}
\label{s:initial_cond}
\textcolor{black}{The initial density profile for the model F\_nW. The density is modeled using Equation~\ref{eq:density_m} with $\eta = 6$ and $\theta_f = 30^\circ$. The color map represents the logarithmic scale of the density $\log_{10} \rho$ in units of $\mathrm{g,cm^{-3}}$. Iso-density contours are overlaid on the plot, shown in black, to highlight the levels of constant density.}
\begin{figure}
    \centering
    \includegraphics[width=\columnwidth]{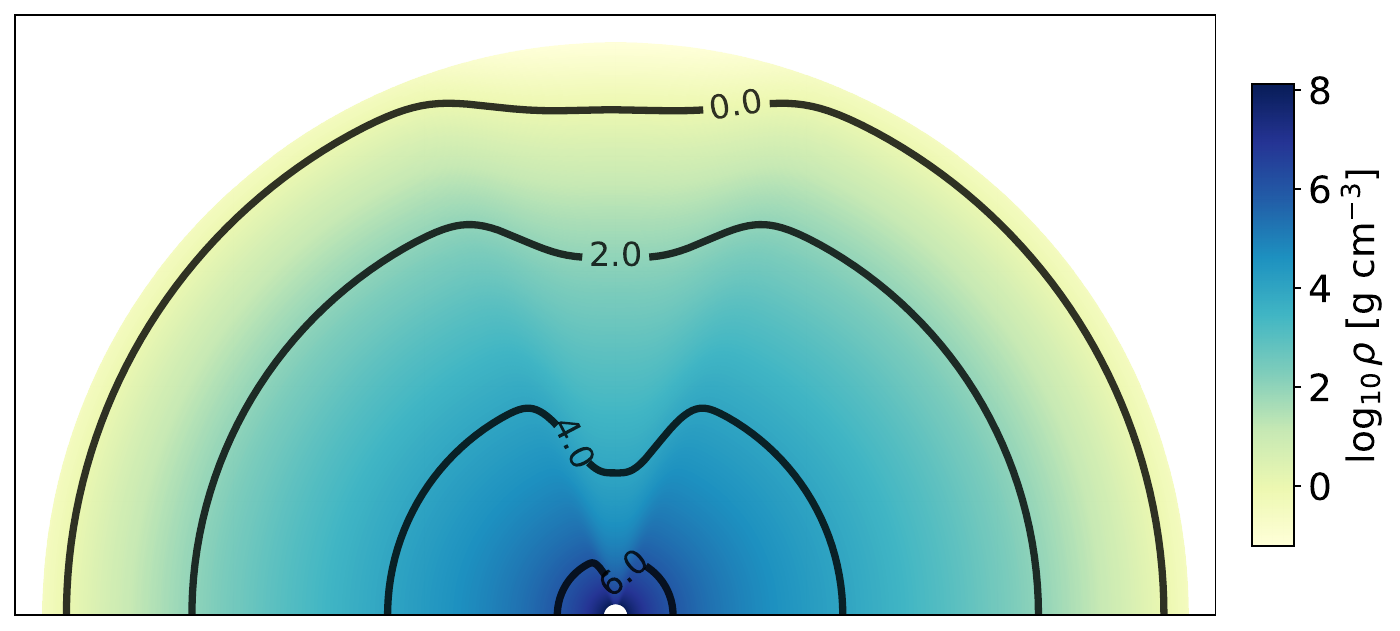}
    \caption{\textcolor{black}{Initial density profile for model F\_nW (model with $\eta = 6$ and $\theta_f = 30^{\circ}$). The 2D color map shows $\log_{10} \rho$ in units of $\mathrm{g\,cm^{-3}}$. The black contours indicate iso-density levels ($\rho = (1, 10^{2}, 10^{4}, 10^{6})\mathrm{g\,cm^{-3}}$).}}
    \label{f:init_cond}
\end{figure}

\section{Jet-Head}
\begin{figure}
    \centering
    \includegraphics[width=\columnwidth]{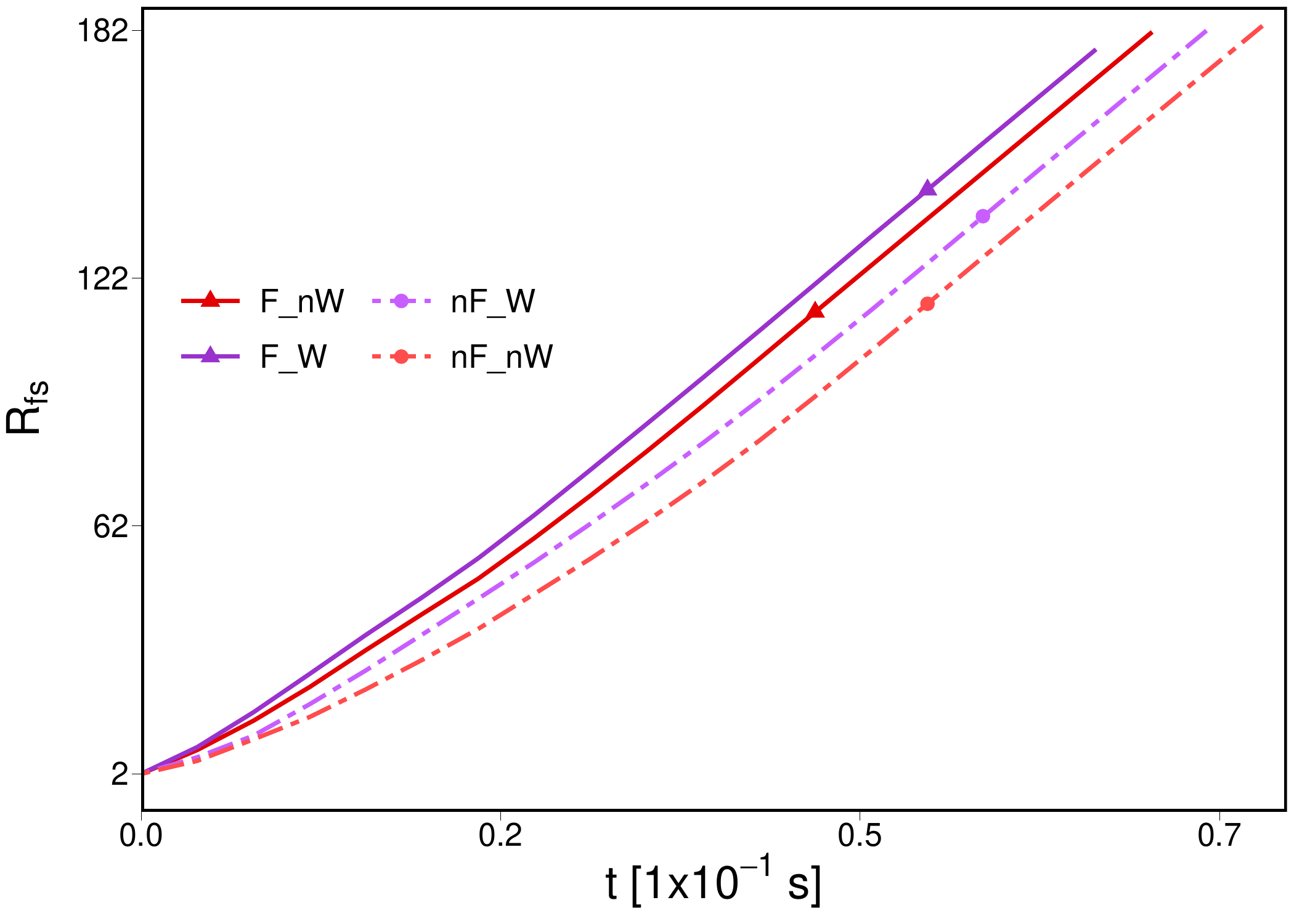}
    \caption{\textcolor{black}{Evolution of the jet head over time for various models (F\_nW: solid red line, F\_W: solid purple line, nF\_W: dashed purple line, and nF\_nW: dashed red line). Dots and triangles represent the breakout time for each model.}}
    \label{f:jet-head}
\end{figure}

\textcolor{black}{Figure \ref{f:jet-head} presents the evolution of the jet head over time for various models (F\_nW: solid red line, F\_W: solid purple line, nF\_W: dashed purple line, and nF\_nW: dashed red line). Dots and triangles mark the breakout time for each model: 0.045~s for F\_nW, 0.05625~s for F\_W, 0.0525~s for nF\_W, and 0.0525~s for nF\_nW. The curves are truncated shortly after breakout, as all models continue to follow the same overall trend beyond this point.} 

\section{Thermal to kinetic energy}
\label{a:Energy_conversion}
\begin{figure}
    \centering
    \includegraphics[width=\columnwidth]{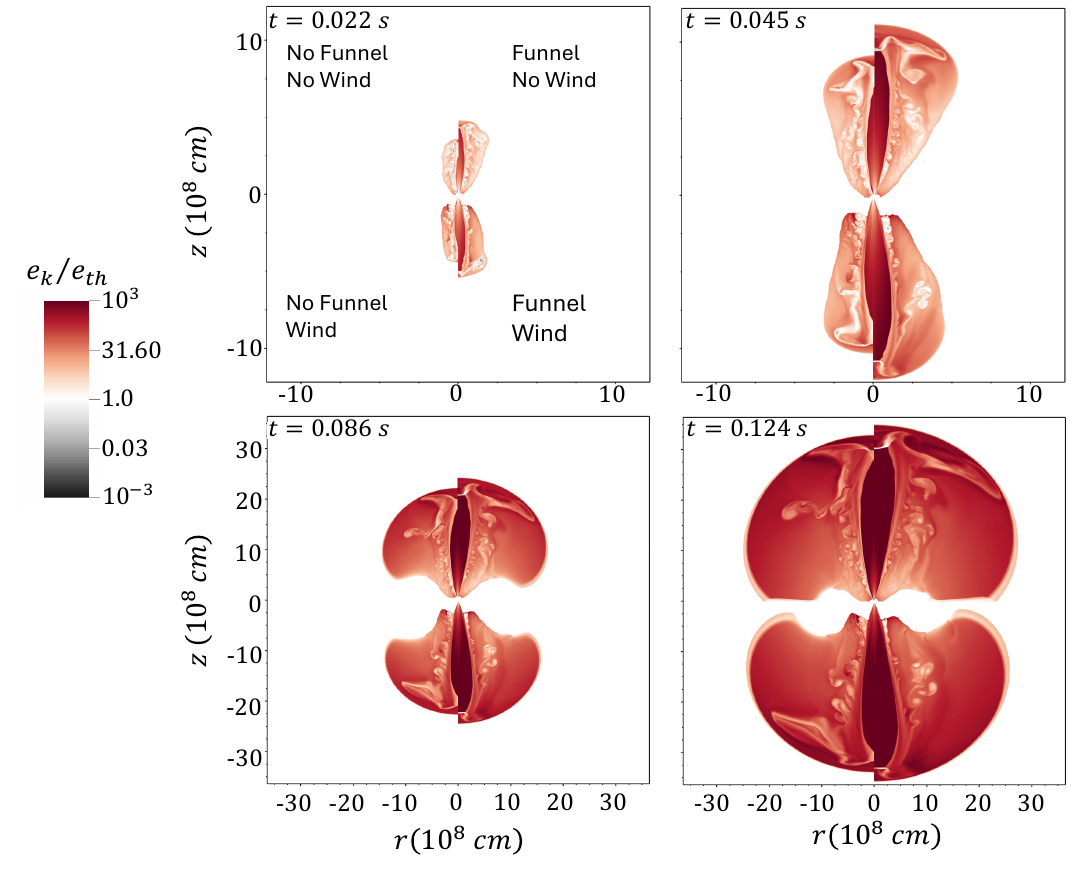}
    \caption{\textcolor{black}{Kinetic-to-thermal energy ratio ($e_k / e_{th}$) for a relativistic jet moving through a static medium without funnel (nF\_nW, upper left), the static medium with funnel (F\_nW, upper right), the expanding medium without funnel (nF\_W, lower left), and the expanding medium with funnel (F\_W, lower right). Four different times are shown ($\rm{t} = 0.022~\mathrm{s}$, $0.045~\mathrm{s}$, $0.086~\mathrm{s}$, and $0.124~\mathrm{s}$). Note that the spatial domain sizes differ between the upper and lower panels.}}
    \label{f:Kinetic_Energy}
\end{figure}

\textcolor{black}{Figure \ref{f:Kinetic_Energy} presents kinetic-to-thermal energy ratio ($e_k / e_{th}$) for a relativistic jet moving through a static medium without funnel (nF\_nW, upper left), the static medium with funnel (F\_nW, upper right), the expanding medium without funnel (nF\_W, lower left). The evolution of the jet is presented at four different times $\rm{t} = 0.022~\mathrm{s}$, $0.045~\mathrm{s}$, $0.086~\mathrm{s}$, and $0.124~\mathrm{s}$.} 

\bsp
\label{lastpage}
\end{document}